\documentclass[twoside]{article}
\usepackage{fleqn,espcrc2}
\usepackage{graphicx}
\usepackage{amsmath}
\usepackage{amsfonts}
\usepackage{amssymb}
\begin{document}

\title{Josephson coupling and plasma resonance in vortex crystal}
\author{A.\ E.\ Koshelev \address
{Materials Science Division, Argonne National Laboratory,
Argonne, IL 60439} and L.\ N.\ Bulaevskii 
\address{Los Alamos National Laboratory, Los Alamos, NM
 87545}\thanks{This work was supported by the NSF Office of the Science and
Technology Center under contract No. DMR-91-20000 and by the U. S. DOE,
BES-Materials Sciences, under contract No. W-31- 109-ENG-38. Work in Los
Alamos is supported by U.S. the DOE.}}
\begin{abstract}
We investigate the magnetic field dependence of the plasma resonance
frequency in vortex crystal state.  We find that low magnetic field
induces a small correction to the plasma frequency proportional to the
field.  The slope of this linear field dependence is directly related
to the average distance between the pancake vortices in the
neighboring layers, the \textit{wandering length}.  This length is
determined by both Josephson and magnetic couplings between layers. 
At higher fields the plasma frequency is suppressed collectively and
is determined by elastic energy of the vortex lattice.  Analyzing
experimental data, we find that (i) the wandering length becomes
comparable with the London penetration depth near T$_{c}$, (ii) at
small melting fields ($< 20$ G) the wandering length does not change
noticeably at the melting transition demonstrating existence of the line
liquid phase in this field range, and (iii) the self consistent theory
of pancake fluctuations describes very well the field dependence of
the Josephson plasma resonance frequency up to the melting point.
\end{abstract}
\maketitle
\section{Introduction}

\begin{figure}[tbh]
\vspace{-.5in}
\includegraphics[width=2.9in]{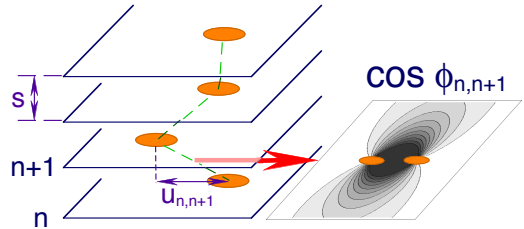} 
\vspace{-.4in}
\caption{Meandering of pancakes along the vortex line in layered
superconductors and grayscale plot of $\cos\phi_{n,n+1}$ near two
misaligned pancakes.}
\label{FigMeand}
\end{figure}

Josephson coupling characterizes the ability of layered superconductors to
carry supercurrents across the layers.  In very anisotropic
superconductors this coupling is suppressed by magnetic field applied
along the c-axis.  Thermal fluctuations and uncorrelated pinning lead
to misalignment of pancake vortices induced by the magnetic field
\cite{gk} (see Fig.\ \ref{FigMeand}).  Misalignment results in nonzero
phase difference and in the suppression of the Josephson interlayer
coupling.  This suppression is quantitatively characterized by the
``local coherence factor'' $\mathcal{C}\equiv\langle\cos\varphi
_{n,n+1}(\mathbf{r})\rangle$, where $\varphi_{n,n+1}(\mathbf{r})$ is
the gauge-invariant phase difference between layers $n$ and $n+1$,
$\langle \ldots\rangle$ means average over thermal disorder and
pinning.  Josephson plasma resonance (JPR) measurements in highly
anisotropic layered superconductors \cite{JPRliq} probe directly the
interlayer Josephson coupling and the effect of pancake vortices on
this coupling, because the squared JPR frequency, $\omega_{p}^{2}$, in
the most part of vortex phase diagram, is proportional to the average
interlayer Josephson energy \cite{bul},
\begin{equation}
\omega_{p}^{2}\approx\omega_{0}^{2}\mathcal{C}\propto J_{0}\mathcal{C},
\label{omegap}
\end{equation}
where $\omega_{0}(T)=c/\sqrt{\epsilon_{0}}\lambda_{c}(T)$ is the zero field
plasma frequency, $\lambda_{c}(T)$ is the c-component of the London
penetration depth, $\epsilon_{0}$ is dielectric constant, and $J_{0}$ is the
Josephson critical current. 
\vspace{-0.35in}

The JPR measurements performed in the liquid vortex phase at relatively high
magnetic fields, $B>B_{J}=\Phi_{0}/\lambda_{J}^{2}$, revealed that the plasma
frequency drops approximately as $1/\sqrt{B}$ \cite{JPRliq}, where $\lambda
_{J}=\gamma s$ is the Josephson length, $\gamma$ is the anisotropy ratio and
$s$ is the interlayer distance. The above dependence is characteristic for the
pancake liquid weakly correlated along the $c$ axis.\cite{kosh}  In this phase
many pancake vortices contribute to the suppression of the phase difference at
a given point. In contrast, in the vortex solid pancake vortices form aligned
stacks and suppression of coupling is caused by weak misalignments of 
the pancake vortices due to the thermal fluctuations and random pinning.
JPR measurements in Bi$_{2}$Sr$_{2}$CaCu$_{2}$O$_{8-\delta}$ (Bi-2212)
crystals \cite{sh,mats} have shown that the JPR frequency decreases
approximately linearly with field in the vortex solid.  In the fields
above 20 Oe the interlayer phase coherence changes drastically at the
transition line, implying the decoupling nature of the first-order
melting transition in agreement with theoretical expectations (see,
e.g., Ref.~\cite{BlatterPRB96}).  On the other hand, at smaller field
phase coherence does not change considerably at the melting point.\cite{sh}

In this paper we consider the Josephson coupling and JPR in the vortex
lattice.  We focus on the suppression of coupling due to thermal
fluctuations of pancake vortices near the equilibrium crystal
positions, and neglect influence of pinning potential.  For real
Bi-2212 crystals this approximation is justified at sufficiently high
temperatures ($\gtrsim 40$ K).  This problem has been considered in
the past in the simple limiting cases.\cite{gk,bulSmB} However
quantitative calculation suitable for comparison with existing JPR
data \cite{sh,mats} in a wide field range is still absent.  At
small fields, when vortices act independently, $\omega_{p}^{2}$
decreases linear with $B$.  The linear dependence was observed
experimentally in Refs.~\cite{sh,mats} in solid state in Bi-2212
crystals.  In fields below 20 Oe near $T_{c}$ this linear dependence
extends to the liquid state providing evidence for a line structure of
the vortices in the liquid at low fields.  The regime of independent
vortices has been considered in Ref.\ \cite{bulSmB}.  In this paper we
extend our consideration to higher fields up to the melting field.

\vspace{-0.1in}
\section{Low fields. Isolated vortex lines}

Consider small magnetic fields $B\ll B_{J}, B_{\lambda}\equiv \Phi_{0}/4\pi\lambda_{ab}^{2},$. 
At these fields regions of suppressed coupling are localized near the
vortex lines (pancake stacks) and do not overlap (single vortex
regime)\cite{bulSmB}.  The field-induced change in $\mathcal{C}$ in
this regime is given by $\delta\mathcal{C}\equiv 1-\mathcal{C}
=BI/\Phi_{0}$, where $I=\int d^{2}\mathbf{r}\left( 1-\cos\left(
\varphi_{n,n+1} (\mathbf{r})\right) \right) $, and
$\varphi_{n,n+1}(\mathbf{r})$ is phase difference induced by
fluctuation displacements $\mathbf{u}_{n}$ in a single line.  The same
integral determines the tilt stiffness due to the Josephson coupling. 
We split integration domain in $I$ into two region, $r<R$ and $r>R$,
where $R$ is the intermediate scale $r_{w}<R<\gamma s$, with
$r_{w}^{2} \equiv\left\langle \mathbf{u}_{n,n+1}^{2}\right\rangle $
and $\mathbf{u} _{n,n+1}\equiv\mathbf{u}_{n+1}-\mathbf{u}_{n}$.  At
$r<R$ we can neglect screening due to the Josephson currents and take
\[
\varphi_{n,n+1}(\mathbf{r})=\arctan\frac{y\! -\! u_{yn+1}}{x\! -\! u_{xn+1}}-\arctan
\frac{y\! -\! u_{yn}}{x\! -\! u_{xn}}.
\]
At $r>R$ we can take $\varphi_{n,n+1}$in linear approximation with
respect to $\mathbf{\ u} _{n}$,
\[
\varphi_{n,n+1}(\mathbf{r},k_{z})\approx\!-\!\int\!\frac{sdk_{z}}{2\pi}\tilde
{k}_{z}\left[  \mathbf{u}(k_{z})\!\times\!\mathbf{\nabla}\right]  \mathrm{K}
_{0}\!\left(\! \frac{\tilde{k}_{z}r}{\gamma}\!\right),
\]
where $\mathbf{u}(k_{z} )\!=\!s\sum_{n}\exp(-isk_{z}n)\mathbf{u}_{n}$,
$\tilde{k}_{z} \!\equiv\!(2/s)\sin(sk_{z}/2)$, and
$\mathrm{K}_{0}\!\left( z\right)$ is the modified Bessel function.  At
intermediate distances $r_{w}\ll r\ll\lambda_{J}$, both expressions
give the same simple result
$\varphi_{n,n+1}(\mathbf{r})\approx\lbrack\mathbf{r}\times\mathbf{u}
_{n,n+1}]/r^{2}$. Using above asymptotics of $\varphi_{n,n+1}$ we obtain
\begin{align*}
I &  =\frac{\pi}{2}\int\frac{dk_{z}}{2\pi}\left(  1-\cos\left(  sk_{z}\right)
\right)  \left|  u(k_{z})\right|^{2}\\
&  \times\ln\left(  \frac{3.72\lambda_{J}^{2}}{u_{n,n+1}^{2}\left(
1-\cos\left(  sk_{z}\right)  \right)  }\right).
\end{align*}
Weak logarithmic dependence on $u_{n,n+1}$ leads to the
nonharmonic tilt energy.\cite{NonLin} In the following we will 
eliminate 
this nonharmonicity using the self consistent harmonic approximation
(SCHA), which results in the substitution $\ln(A/u_{n,n+1}^{2})
\rightarrow\ln(0.24A/r_{w}^{2})$.  Approximate evaluation of the above
integral gives a simple practical relation connecting the field-induced
correction of the plasma frequency $\omega_{p}(B,T)$ with $r_{w}$ for
the case $r_{w}<\lambda_{J}<a$:
\begin{equation}
\frac{\omega_{0}^{2}(T)-\omega_{p}^{2}(B,T)}{\omega_{0}^{2}(T)}\approx
\frac{\pi Br_{w}^{2}}{2\Phi_{0}}\ln\frac{0.8\lambda_{J}}{r_{w}}.
\label{Cosrw}
\end{equation}
This relation allows one to extract $r_{w}^{2}$ from the plasma
resonance measurements.

We now calculate $r_{w}^{2}$ when wandering of the vortex lines is caused by
thermal fluctuations. In the single vortex regime $r_{w}^{2}$ is determined by
the wandering energy consisting of the Josephson and magnetic contributions,
\begin{equation}
\mathcal{F}_{w}\approx\frac{1}{2}\int\frac{dk_{z}}{2\pi}\left[  \varepsilon
_{1J}\tilde{k}_{z}^{2}+w_{M}\right]  \left|  u(k_{z})\right|  ^{2},
\end{equation}
where $\varepsilon_{1J}\approx(\varepsilon_{0}/\gamma^{2})\ln\left(
1.33\gamma/(r_{w}\tilde{k}_{z})\right) $ is the line tension due to
the Josephson coupling, $w_{M}\approx(\varepsilon_{0}/\lambda_{ab}^{2}
)\ln(1.5\lambda_{ab}/r_{w})$ is the effective cage potential, which
appears due to nonlocal magnetic interactions between pancake vortices
in different layers (it describes the magnetic tilt stiffness at wave
vectors $k_{z}>1/r_{w}$)\cite{kkPRB93,BlatterPRB96}, and
$\varepsilon_{0}\equiv \Phi_{0}^{2}/(4\pi\lambda_{ab})^{2}$.  Assuming
Gaussian fluctuations of pancake vortices we obtain
\begin{equation}
r_{wT}^{2}=\frac{8T}{sw_{M}}\frac{1}{1+\zeta+\sqrt{1+\zeta}}, \label{rwT}
\end{equation}
where the parameter $\zeta(T)\approx4\lambda_{ab}^{2}(T)/\lambda_{J}^{2}$
describes the relative roles of the Josephson and magnetic interactions.
Substituting this result into Eq.~(\ref{Cosrw}) we obtain
\begin{equation}
\frac{\omega_{0}^{2}-\omega_{p}^{2}(B)}{\omega_{0}^{2}}\approx
\frac{4\pi\lambda_{ab}^{2}BT}{s\epsilon_{0}\Phi_{0}}
\frac{1}{1+\zeta+\sqrt{1+\zeta}}.
\label{SmBTheor}
\end{equation}
This result of the single vortex regime is valid in both solid and
liquid states for $B\ll B_{J}$, because in this field range
wandering of lines at short scales does not change much at the melting
point.  Eq.\ (\ref{SmBTheor}) describes fairly well the suppression of
the plasma frequency at small fields \cite{bulSmB}.  With the data of Ref.\
\cite{sh} for underdoped Bi-2212 with T$_{c}\approx 84.5$ K 
Eq.\ (\ref{Cosrw}) gives unexpectedly large wandering length
$r_{w}\approx 1$$\mu$m at $77$ K, which is comparable with both
$\lambda_{ab}$ and $\lambda_{J}$ at this temperature.  However, we
found that this estimate is a in good agreement with the theoretical
calculation (\ref{rwT}).

\vspace{-0.1in}
\section{High fields, $B>B_{J}$, $B_{\lambda}$}

As the field increases two competing effects start to influence
pancake fluctuations and field dependence of the average Josephson
energy.  Vortex interactions diminish pancake fluctuations.  On the
other hand, collective suppression of the Josephson energy decreases
tilt stiffness and enhances pancake fluctuations.  Using general
relations connecting the phase perturbations with the elastic lattice
deformations (see, e.g. Ref.~\cite{gk}), we obtain
\begin{align}
&  \!\!\delta\mathcal{C}
=\frac{\left\langle \left[  \varphi_{n,n+1}\right]^{2}\right\rangle }{2}
\approx\frac{\left(  2\pi sn_{v}\right)  ^{2}}
{2}\label{DeltaChighB}\\
&  \!\!\!\times\!\!\int\!\!\frac{d^{2}\!qdk_{z}}{\left(  2\pi\right)  ^{3}}
\!\!\sum\limits_{Q<q_{m}}\!\!\!\frac{\tilde{k}_{z}^{2}\!\left(  \left[
\mathbf{Q}\!\times\!\mathbf{q}\right]  ^{2}\!\langle\mathbf{u}_{l}^{2}
\rangle\!+\!\left(  q^{2}\!+\!\mathbf{Qq}\right)  ^{2}\!\langle\mathbf{u}
_{t}^{2}\rangle\right)  }{q^{2}\left(  \left(  \mathbf{q}+\mathbf{Q}\right)
^{2}+\tilde{k}_{z}^{2}/\gamma^{2}\right)  ^{2}},\nonumber
\end{align}
where $n_{v}\equiv B/\Phi_{0}$, $\mathbf{Q}$ are the reciprocal lattice vectors (the cut off 
$q_{m}\approx 2.2/r_{w}$ in the summation over $\mathbf{Q}$ is established by 
comparison with the single vortex regime), and $\mathbf{u}
_{l}(\mathbf{q},k_{z})$ ($\mathbf{u}_{t}(\mathbf{q},k_{z})$) are the
longitudinal (transversal) elastic lattice displacements.  Integration
with respect to the in-plane wave vector $\mathbf{q}$ is limited by
the first Brillouin zone, which we approximate by the circle
$q<K_{0}$, $K_{0}^{2}=4\pi n_{v}$.  For thermal Gaussian
fluctuations the mean squared averages $\langle
\mathbf{u}_{t,l}^{2}\rangle\equiv \langle
|\mathbf{u}_{t,l}(\mathbf{q},k_{z})|^{2}\rangle$ are determined by the corresponding
components of the elastic matrix $\Phi_{t,l}(\mathbf{q},k_{z})$
\[
\langle|\mathbf{u}_{t,l}(\mathbf{q},k_{z})|^{2}\rangle 
=T/\Phi_{t,l}(\mathbf{q},k_{z}).
\]
At high fields $B>B_{J}$, $B_{\lambda}$, and
at large $k_{z}$, $k_{z}\gg1/\lambda_{ab}$, $\sqrt{n_{v}}$ we have 
\begin{align*}
\Phi_{t}(q,k_{z})&=C_{66}q^{2}+\Phi_{44}(q,k_{z}),\\
\Phi_{l}(q,k_{z})&  =\Phi_{11}(q)+\Phi_{44}(q,k_{z}),
\end{align*}
where $\Phi_{11} \approx\frac{B^{2}}{4\pi}\left(
1-\frac{q^{2}}{4K_{0}^{2}}\right)$ is the compression stiffness,
$C_{66}=A_{66}n_{v}\epsilon_{0}/4$ is the shear modulus,  parameter
$A_{66}<1$ describes fluctuation suppression of $C_{66}$, which we
approximate as $A_{66}=1-0.4B/B_{m}$ with $B_{m}$ being the melting
field, and
\begin{align*}
&  \Phi_{44}(q,k_{z})\approx\frac{n_{v}\varepsilon_{0}k_{z}^{2}
}{2\gamma^{2}}\ln\!\frac{0.11a^{2} }{r_{w}^{2}\left(  1\!-\!0.53q^{2}
\!/\!K_{0}^{2}\right)  ^{2}}\\
&  \!+\frac{n_{v}\varepsilon_{0}}{2\lambda_{ab}^{2}}\ln\!\left(
0.5+\frac{0.13a^{2}}{r_{w}^{2}}\right)  +\frac{B^{2}} {4\pi\lambda_{ab}^{2}
}\frac{k_{z}^{2}}{k_{z}^{2}\!+\!\gamma^{2}q^{2}},
\end{align*}
is the tilt stiffness, computed within SCHA, with $a\equiv 
1/\sqrt{n_{v}}$. The wandering length $r_{w}$ has to be determined 
self-consistently  from the  equation
\begin{eqnarray}
r_{w}^{2}&=&2\!\int\!\!\frac{d^{2}\!qdk_{z}}{\left( 2\pi\right)^{3}}
\left(1-\cos(sk_{z})\right) \label{WanLen}\\
&\times&\left(\langle|\mathbf{u}_{l}(\mathbf{q},k_{z})|^{2}\rangle +
\langle|\mathbf{u}_{t}(\mathbf{q},k_{z})|^{2}\rangle\right).\nonumber
\end{eqnarray}
Expression (\ref{DeltaChighB}) for $\delta\mathcal{C}$ can be naturally split 
into the collective contribution $\delta\mathcal{C}_{coll}$, corresponding 
to $Q=0$ term in the $Q$-summation, 
\begin{equation}
\!\!\delta\mathcal{C}_{coll}\approx\frac{\left(2\pi sn_{v}\right)  ^{2}}{2}
\!\!\int\!\!\frac{d^{2}\!qdk_{z}}{\left( 2\pi\right)^{3}}
\!\frac{\tilde{k}_{z}^{2}  q^{2}\!\langle\mathbf{u}_{t}^{2}\rangle}
{\left(\mathbf{q}^{2}+\tilde{k}_{z}^{2}/\gamma^{2}\right) ^{2}},
\label{Coll}
\end{equation}
and the local contribution $\delta\mathcal{C}_{loc}$ coming from $Q>0$
terms.  
At high fields $B\gg \Phi_{0}/\lambda_{J}^{2}$ we obtain approximate expression for 
$\delta\mathcal{C}_{loc}$, which resembles the single-vortex result 
(\ref{Cosrw})
\begin{equation}
\!\!\delta\mathcal{C}_{loc}\approx\frac{\pi 
n_{v}r_{w}^{2}}{2}\ln\frac{0.58a}{r_{w}}.
\label{CollTerm}
\end{equation}
If we consider suppression of coupling in the cylindrical Bravais cell
near the chosen vortex line, then the local term determines suppression of
coupling caused by this vortex line and the collective term determines
suppression of coupling caused by all other vortex lines.  In general,
relative role of the collective term in $\delta\mathcal{C}$ grows with
field.  We use above expressions to calculate the field dependence of
$\mathcal{C}$ for comparison with JPR data.

\begin{figure}[tbh]
\vspace{-.3in} 
\includegraphics[width=2.8in]{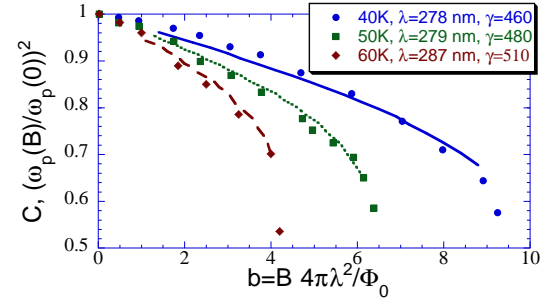} 
\vspace{-.3in} \caption{Comparison of the field dependence of the
reduced plasma frequency squared (courtesy of M.\ Gaifullin and Y.\
Matsuda \protect\cite{mats}) and calculated field dependence of the
local coherence factor $\mathcal{C}$. }
\label{FigCB}
\end{figure}

Recently detailed measurements of field dependence of JPR frequency in
the vortex crystal state of Bi-2212 have been done by M.\ Gaifullin
\textit{et al.} using frequency scan.\cite{mats} To compare our
calculations with JPR data we need to know $\lambda_{ab}(T)$ and
$\gamma=\lambda_{c} /\lambda_{ab}$.  $\lambda_c$ is extracted directly
from JPR frequency at $B=0$,
$\lambda_{c}(T)=c/\sqrt{\epsilon_0}\omega_0(T)$ taking $\epsilon_0=11$
and $\gamma$ is chosen as a fitting parameter.  Fig. \ref{FigCB}
compares the computed dependence $\mathcal{C}(B)$ with 
$(\omega_p(B)/\omega_p(0))^2$ for
three values of temperature.  We also show obtained values of
$\lambda\equiv \lambda_{ab}$ and $\gamma$.  We obtain $\gamma$, that slightly grows
with temperature (from $460$ at $40$ K to $510$ at $60$ K).  Such
enhancement of $\gamma $
is expected due to the phase fluctuations.

In conclusion, we have calculated the field dependence of the JPR
frequency in the vortex crystal.  In the single vortex regime at low
magnetic fields the JPR provides a direct probe for meandering of
individual lines.  The wandering length extracted from the JPR data is in
agreement with the theoretical calculations.  Our theory of
pancake fluctuations gives a very good description of the field
dependence of the plasma frequency up to the melting field.  The
authors thank M.\ Gaifullin, Y.\ Matsuda, T.\ Tamegai, and T.\
Shibauchi for providing their experimental data prior publication and
V.\ Vinokur for constructive comments.


\vspace{-.4in} 
\begin{thebibliography}{99}
\bibitem{gk}L.\ I.\ Glazman and A.\ E.\ Koshelev, Phys. Rev. B, \textbf{43}
 (1991) 2835; L. Daemen \textit{et al.},
Phys.\ Rev.\ Lett.\ \textbf{70}, 1167 (1993); Phys.\ Rev.\   B 
\textbf{47} (1993) 11291;

\bibitem {JPRliq}Y.\ Matsuda \textit{et al.},
Phys.\ Rev.\ Lett.
\textbf{75} (1995) 4512;  \textit{ibid} \textbf{78} (1997) 1972;
O.~K.~C.~Tsui \textit{et al.},
Phys.~Rev.~Lett., {\bf 73} (1994) 724; 
\textit{ibid} {\bf 76} (1996) 819; 
N.~P.~Ong \textit{et al.},
Physica C \textbf {293} (1997) 20.

\bibitem {bul}L.\ N.\ Bulaevskii, M.\ P.\ Maley, and M.\ Tachiki,
Phys.\ Rev.\ Lett. \textbf{74} (1995) 801 .

\bibitem {kosh}A.\ E.\ Koshelev, Phys.\ Rev.\ Lett. \textbf{77} (1996) 3901;
A.\ E.\ Koshelev, L.\ N.\ Bulaevskii, and M.\ P.\ Maley, Phys. Rev. Lett.
\textbf{81} (1998) 9.
%
%

\bibitem {sh}T.\ Shibauchi \textit{et al.},
Phys.\ Rev.\ Lett.\ \textbf{83} (1999) 1010;
%

\bibitem {mats}M.\ Gaifullin \textit{et al.}, preprint.

\bibitem {BlatterPRB96}G.\ Blatter \textit{et al.}, Phys.\ Rev.\ B
\textbf{54} (1996) 72.

\bibitem{bulSmB}L.\ N.\ Bulaevskii \textit{et al.},
cond-mat/9907462

\bibitem{NonLin}C.\ Kr\"{a}mer, Physica C \textbf{256} (1996) 236;
T.\ R.\ Goldin and B.\ Horovitz, Phys.\ Rev.\ B
\textbf{58} (1998) 9524.
%
%

\bibitem {kkPRB93}A.\ E.\ Koshelev and P.\ H.\ Kes, Phys.\ Rev.\ B
\textbf{48} (1993) 6539.
%
%

\end{thebibliography}
\end{document}